\documentclass[twocolumn,showpacs,preprintnumbers,amsmath,amssymb]{revtex4}
\usepackage{graphicx}% Include figure files
\usepackage{dcolumn}% Align table columns on decimal point
\usepackage{bm}% bold math
\usepackage{epsfig}
\begin{document}

%\preprint{APS/123-QED}

\title{Retardation and flow at the glass transition}% Force line breaks with \\

\author{U. Buchenau}
 \email{buchenau-juelich@t-online.de}
\affiliation{%
J\"ulich Center for Neutron Science, Forschungszentrum J\"ulich\\
Postfach 1913, D--52425 J\"ulich, Federal Republic of Germany
}%
\date{February 9, 2016}% It is always \today, today,

\begin{abstract}
The crossover from back-and-forth jumps between structural minima to the no-return jumps of the viscous flow is modeled in terms of an ensemble of double-well potentials with a finite decay probability. The ensemble is characterized by the Kohlrausch-exponent $\beta$ of the time dependence $t^\beta$ of the response at short times. The model is applied to shear and dielectric data from the literature.
\end{abstract}

\pacs{78.35.+c, 63.50.Lm}% PACS, the Physics and Astronomy
\maketitle

\section{Introduction}

The energy landscape of a deeply undercooled liquid seems to be well described within the concept of inherent states \cite{palmer,stillinger,scio1,scio2,heuer}. An inherent state is a structurally stable energy minimum configuration of the particles of the liquid. The passage from one inherent state to another seems to occur by thermally activated jumps. In spite of this reasonably well-understood basis, the flow process in undercooled liquids is still the subject of conflicting interpretations \cite{cavagna,ngai,nemilov,dyre}. 

Consider a single jump between two inherent states of the glass or the liquid. This jump must be a local event, because otherwise it would have an infinite energy barrier. One can describe the jump as a structural rearrangement of a finite core, a group of $N$ neighboring atoms or molecules. The change of the volume and of the shape of the core determine the coupling of the local jump to the elastic constants in Eshelby's classical picture \cite{eshelby}. 

If the lifetime $\tau_c$ of the local structure is long compared to the relaxation time of the jump, there will be many back-and-forth jumps within $\tau_c$. The question is: How much of the response is due to these back-and-forth jumps (the retardation response \cite{ferry}) and how much is due to the final viscous flow? A second important question is: How sharp is $\tau_c$? In other words: Do all local structures decay with the same time constant? In spite of all our theoretical and numerical work \cite{cavagna}, we do not yet have clear answers to these questions.

The present paper intends to contribute to this answer in terms of a simple pragmatical model, an ensemble of double-well potentials with different relaxation times $\tau_r$ and a common lifetime $\tau_c$, possibly a broadened one. The model is fitted to a large amount of shear and dielectric data from the literature.

The paper is organized as follows: After this introduction, Section II describes the model. Section III contains the comparison to experiment. Section IV discusses the results and concludes the paper.

\section{The model}

\subsection{Shear}    

The model describes an ensemble of structural rearrangements in an undercooled liquid. The structural rearrangement is supposed to occur within an inner core of $N$ atoms or molecules. The distortion of the core couples to the surrounding elastic matrix according to the Eshelby mechanism \cite{eshelby}. 

In the Eshelby theory, the structural jump of the central core couples to the stress, not to the strain \cite{eshelby}. This implies that the effects of different structural jumps do not add in the shear modulus, but in the elastic shear compliance \cite{ferry}
\begin{equation}\label{jom}
J(\omega)=\frac{1}{G}+\int_{-\infty}^\infty \frac{L(\tau)}{1+i\omega\tau}d\ln\tau-\frac{i}{\omega\eta}.
\end{equation}

Here $G$ is the infinite frequency shear modulus and $\eta$ is the viscosity. $L(\tau)$ describes the density of the retardation processes (reversible relaxation processes \cite{ferry}) at the relaxation time $\tau$. 

A third material constant hidden in this equation is the zero-frequency recoverable compliance $J_0$, the infinite frequency elastic compliance plus the integral over the retardation processes
\begin{equation}\label{j0e}
	J_0=\frac{1}{G}+\int_{-\infty}^\infty L(\tau)d\ln\tau.
\end{equation}

Eq. (\ref{jom}) makes a separation of two independent relaxation contributions to the compliance, the retardation spectrum and the viscosity. The retardation spectrum is due to back-jumps into the initial inherent state, the viscosity is due to no-return processes. The crossover from the back-jumps to the no-return jumps occurs at a cutoff relaxation time $\tau_c$, the lifetime of the local structure under the influence of jumps in the neighborhood.

The model assumes an ensemble of double-well potentials with different relaxation rates $r=1/\tau_r$, which all decay with the same rate (or approximately the same rate) $r_c=1/\tau_c$. Each mode supplies a viscous contribution to the terminal decay at $\tau_c$ and a retardation or back-jump contribution at the relaxation time
\begin{equation}\label{tau}
\tau=\frac{1}{r+r_c}=\frac{\tau_r\tau_c}{\tau_r+\tau_c}	
\end{equation}
which is always shorter than both $\tau_r$ and $\tau_c$. 

The retardation contribution is weakened by the decay factor $r/(r+r_c)$, because a fraction $r_c/(r+r_c)$ of the jumps between the two sites goes into the terminal decay. Let $l(\tau_r)$ be the relaxation time density with respect to $\tau_r=1/r$. Then the relation
\begin{equation}\label{norm}
	L(\tau)d\ln{\tau}=\frac{r}{r+r_c}\frac{l(\tau_r)}{G}d\ln{\tau_r},
\end{equation}
enables the calculation of $J(\omega)$ from $l(\tau_r)$, $G$ and $\tau_c$.

%%%%%%%%%%%%%%%%%%%%% begin figure %%%%%%%%%%%%%%%%%%%%%%%%%%%%%%%%%%%%%
\begin{figure} 
\hspace{-0cm} \vspace{0cm} \epsfig{file=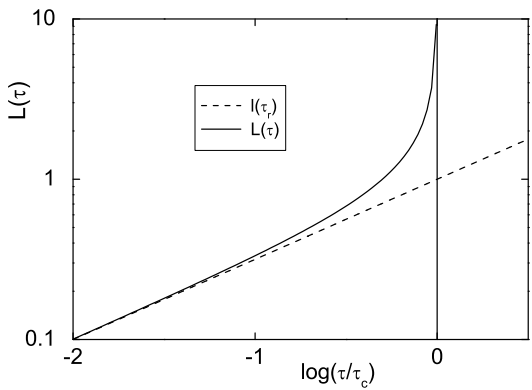,width=7 cm,angle=0} \vspace{0cm} \caption{Comparison of the two functions $L(\tau)$ and $l(\tau_r)$ for $\beta=1/2$.}
\end{figure}
%%%%%%%%%%%%%%%%%%%%% end figure %%%%%%%%%%%%%%%%%%%%%%%%%%%%%%%%%%%%%%%

Note that this definition allows to choose the function $l(\tau_r)=l_0\tau_r^\beta$ with a Kohlrausch exponent $\beta<1$, because the decay prefactor $r/(r+r_c)$ cancels the divergence of the function at large $\tau_r$. In this case
\begin{equation}
	L(\tau)=\frac{l_0}{G}\left(\frac{\tau\tau_c}{\tau_c-\tau}\right)^\beta.
\end{equation}

Fig. 1 shows $L(\tau)$ and $l(\tau_r)$ for $\beta=1/2$ in the neighborhood of $\tau_c$. For small $\tau$, $L(\tau)$ has the same $\tau^\beta$-behavior as $l(\tau_r)$, but diverges as $\tau$ approaches $\tau_c$. Roughly speaking, $L(\tau)$ can be viewed as a sum of a Kohlrausch $\tau^\beta$-function terminating at $\tau_c$, plus a $\delta$-function at $\tau_c$, slightly broadened and shifted to lower $\tau$. For $\beta=1/2$, the Kohlrausch part contains 2/$\pi$ of the total contribution.

The data of the next section will enforce a heterogeneous distribution of different values $\tau_f$ for $\tau_c$ at different places in the sample. Let us postulate a gaussian distribution in $\ln{\tau_f}$ around the average $\tau_c$
\begin{equation}
	\Phi(\tau_f)=\sqrt{\frac{4\ln{2}}{\pi W_r^2}}\exp\left(\frac{4\ln{2}}{W_r^2}(\ln{\tau_f}-\ln{\tau_c})^2\right)
\end{equation}
and let all these different $\tau_f$ values have the same $l(\tau_r)$. Then $L(\tau)$ must be obtained by integrating all contributions from eq. (\ref{norm}) over $\tau_f$.

Note that $W_r=\ln{10}=2.303$ implies a full width at half maximum of a decade in $\tau_f$ around $\tau_c$.

With this last definition, the fit of shear data requires six parameters: The shear modulus $G$, the recoverable compliance $J_0$, the viscosity $\eta$, the average decay time $\tau_c$, the width $W_r$ of its distribution, and finally the Kohlrausch parameter $\beta$. $J_0$ determines the dimensionless product $GJ_0$, a measure of the decrease of the elastic response by the retardation processes. For convenience, the viscosity $\eta$ will be reported in terms of the Maxwell time $\tau_M=\eta/G$.

\subsection{Dielectrics and others}

In the case of a molecular liquid, the change of the orientation of the molecules in the core and in its surroundings determines the contribution of the jump to the dielectric susceptibility. The picture is compatible with NMR findings \cite{bohmer} of a bimodal distribution with many small angle orientational jumps and a few large angle ones: the few large angle ones occur within the rearranging core, the many small angle ones in the elastically distorted surroundings.

To apply the concept of structural rearrangements to dielectric data (or to any other dynamic susceptibility influenced by the flow process), one has to calculate the normalized dielectric susceptibility
\begin{equation}
	\Phi(\omega)=\frac{\epsilon(\omega)-\epsilon_\infty}{\epsilon_s-\epsilon_\infty},
\end{equation}
where $\epsilon_s$ is the zero frequency susceptibility and $\epsilon_\infty$ is the high frequency limit.

The dielectric signal consists of a viscous contribution and the retardation spectrum. The comparison to experiment in the next section shows that the terminal viscous contribution does usually appear at $\tau_c$. However, in the mono-alcohols it appears at a much longer terminal relaxation time $\tau_\eta$. 

Each retardation contribution is weakened by the decay factor $r/(r+r_c)$, because a fraction $r_c/(r+r_c)$ of the jumps between the two sites goes into the terminal decay. The two decay channels - inter-site jumps and terminal decay - together must bring the dielectric polarization down to zero. Let $l(\tau_r)$ be the relaxation time density with respect to $\tau_r=1/r$. Normalizing the retardation contribution to 1
\begin{equation}\label{norme}
	\int_{-\infty}^\infty\frac{r}{r+r_c}l(\tau_r)d\ln{\tau_r}=1
\end{equation}
one has the viscosity contribution
\begin{equation}
	\Delta_\eta=\int_{-\infty}^\infty\frac{rr_c}{(r+r_c)^2}l(\tau_r)d\ln{\tau_r}.
\end{equation}
For the Kohlrausch case with $\beta=1/2$, $\Delta_\eta=1/2$, so normally one has two thirds retardation and one third viscous response. 

This is dramatically different in the case of the mono-alcohols, where the terminal relaxation time $\tau_\eta$ is much larger than $\tau_c$. Within $\tau_\eta$, a number $n_r=\tau_\eta/\tau_c$ of generations of retardation processes come and go. So the total viscous response is $n_r\Delta_\eta$ for a retardation response of 1. The total response has the viscosity and retardation fractions
\begin{equation}\label{mono}
	f_\eta=\frac{n_r\Delta_\eta}{1+n_r\Delta_\eta}\ \ f_r=\frac{1}{1+n_r\Delta_\eta}.
\end{equation}
For the usual case $n_r=1$ one returns to values close to 1/3 and 2/3, respectively, but in the mono-alcohols, the reardation retains only a few percent of the total response.

Since one has to reckon with a whole distribution of local structure lifetimes $\tau_c$, one also has to reckon with a broadening of the terminal viscous component at $\tau_\eta$.  One has to replace $1/1+i\omega\tau_\eta$ by the broadened normalized function $A_\eta(\omega)$
\begin{equation}\label{aeta}
	A_\eta(\omega)=\int_{-\infty}^\infty dx  \sqrt{\frac{4\ln{2}}{\pi W_\eta^2}}\frac{\exp(-4\ln{2}x^2/W_\eta^2)}{1+i\omega\exp(x)\tau_\eta},
\end{equation}
where the full width at half maximum $W_\eta$ is again a free parameter.

The normalized dielectric relaxation function $\Phi(\omega)$ is then given by
\begin{equation}\label{phiom}
\Phi(\omega)=f_\eta A_\eta(\omega)+f_r\int_{-\infty}^\infty\frac{r}{r+r_c}\frac{l(\tau_r)}{1+i\omega\tau}d\ln{\tau_r}.
\end{equation}

\section{Comparison to experiment}

\subsection{Shear data}

%%%%%%%%%%%%%%%%%%%%% begin figure %%%%%%%%%%%%%%%%%%%%%%%%%%%%%%%%%%%%%
\begin{figure}
\hspace{-0cm} \vspace{0cm} \epsfig{file=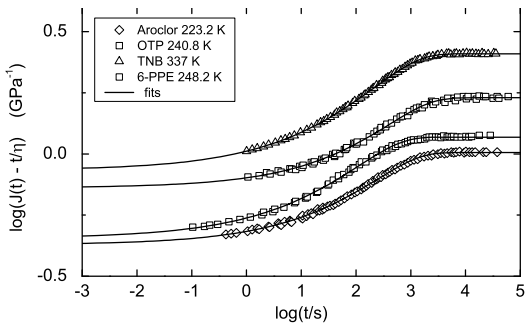,width=7 cm,angle=0} \vspace{0cm} \caption{Recoverable compliance data \cite{plazek-magill,plazek-bero} fitted in terms of an ensemble of relaxators with a distribution of finite lifetimes  (fit parameters see Table I).}
\end{figure}
%%%%%%%%%%%%%%%%%%%%% end figure %%%%%%%%%%%%%%%%%%%%%%%%%%%%%%%%%%%%%%%

If one knows the shear modulus $G$, the viscosity $\eta$, the recoverable compliance $J_0$, the crossover relaxation time $\tau_c$, the width $W_r$ of its distribution and the Kohlrausch parameter $\beta$, one can calculate $J(\omega)$ from eq. (\ref{jom}), determining $L(\tau)$ from eq. (\ref{norm}). Then $G(\omega)=1/J(\omega)$.

Measurements of the time-dependent compliance $J(t)$ can be calculated from \cite{ferry}
\begin{equation}\label{jt}
	J(t)=\int_{-\infty}^\infty(1-\exp(-t/\tau)L(\tau)d\ln{\tau}+\frac{t}{\eta}.
\end{equation}
This equation can be used to fit the recoverable compliance measurements of Plazek and coworkers \cite{plazek-magill,plazek-bero} shown in Fig. 1. 

Plazeks method allows to measure the time dependence of the recoverable part of the compliance, $J_r(t)=J(t)-t/\eta$. In this quantity, the crossover time $\tau_c$ appears in the saturation of $J_r(t)$ at long times, in principle much better visible than in $G(\omega)$-measurements, where the strong viscous response overshadows the saturation of the retardation response. A larger distribution width $W_r$ should lead to a smaller curvature of $J_r(t)$ in the saturation region. The data fix $\tau_c$ within ten to twenty percent, but are unfortunately not accurate enough to decide with final clarity whether $\tau_c$ is sharp or not (see Table I). 

%%%%%%%%%%%%%%%%%%%%% begin figure %%%%%%%%%%%%%%%%%%%%%%%%%%%%%%%%%%%%%
\begin{figure}   
\hspace{-0cm} \vspace{0cm} \epsfig{file=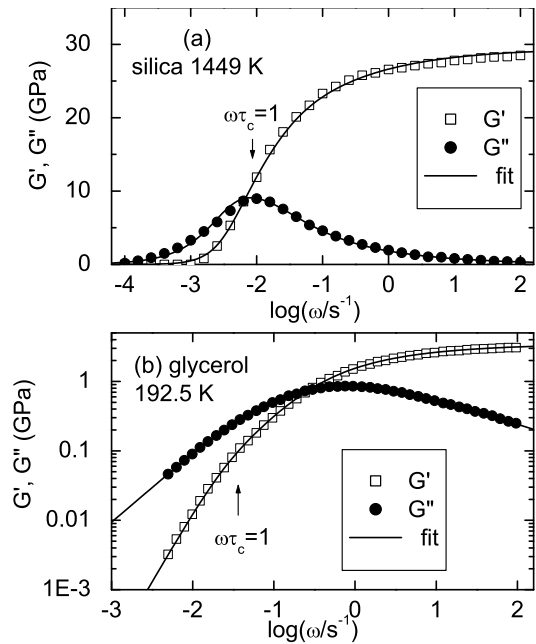,width=7 cm,angle=0} \vspace{0cm} \caption{Fit of dynamical shear data close to the glass transition in terms of the model explained in the text for (a) vitreous silica \cite{mills} (b) glycerol \cite{donth1} (fit parameters see Table I).}
\end{figure}
%%%%%%%%%%%%%%%%%%%%% end figure %%%%%%%%%%%%%%%%%%%%%%%%%%%%%%%%%%%%%%%

\begin{table}[htbp]
	\centering
		\begin{tabular}{|c|c|c|c|c|c|c|c|}
\hline
substance                  & $T$  & $G$   & $GJ_0$  & $\beta$  & $\tau_c$        & $\tau_M$ & $W_r$     \\
\hline   
                            & $K$  & $GPa$ &        &          &     $s$         &   $s$   &            \\
\hline                                                                                           
TNB \cite{plazek-magill}    & 337  & 1.16  & 3.0    &  0.35    & 1362            &         & 0.7$\pm$1.3\\
aroclor \cite{plazek-bero}  & 232.8& 2.30  & 2.3    &  0.38    & 1188            &         & 1.7$\pm$1.7\\
OTP \cite{plazek-bero}      & 240.8& 2.11  & 2.5    &  0.41    & 495             &         & 2.3$\pm$1.3\\
6-PPE \cite{plazek-bero}    & 248.2& 1.32  & 2.3    &  0.44    & 1636            &         & 2$\pm$2   \\
\hline
silica \cite{mills}         & 1449 & 29.5  & 1.7    &  0.36    & 130             &  101    &       0    \\
5-PPE \cite{tina}           & 250  & 1.03  & 2.7    &  0.48    & 4.8             &  0.70   & 3.4$\pm$1.5\\
DC704 \cite{tina}           & 216  & 1.05  & 2.7    &  0.45    & 0.90            &  0.16   & 2.6$\pm$0.9\\
PC \cite{donth2}            & 159  & 0.66  & 3.1    &  0.37    & 14.0            &  0.52   &       1.65*\\
TPE \cite{niss}             & 258  & 1.16  & 2.2    &  0.53    & 0.55*           &  0.17   & 3.0*       \\
DGEBA \cite{donth2}         & 256.3& 0.93  & 2.6    &  0.57    & 11.3*           &  4.0    &       4.1* \\
DEP$^a$ \cite{maggi}        & 183  & 1.36  & 3.5    &  0.43    & 50.7*           &  7.0    & 1.73*      \\ 
\hline		
glycerol \cite{donth1}      & 192.5& 3.48  & 5.9    &  0.43    & 33.4            &  2.7    & 0          \\
PG \cite{maggi}             & 180  & 3.32  & 6.2    &  0.46    & 0.101*          &  0.005  & 1.37*      \\
\hline

		\end{tabular}
	\caption{Fit parameters for dynamical shear data in thirteen glass formers. TNB is tri-naphtyl benzene, OTP is ortho-terphenyl, 6-PPE is 6-polyphenylether, DGEBA is an epoxy resin, PC is propylene carbonate, DC704 is a vacuum pump oil, 5-PPE is 5-polyphenylether, TPE is triphenylethylene, DEP is diethyl phthalate; it is fitted$^a$ with a secondary $\beta$-peak with the amplitude $f_\beta=0.009$. PG is propylene glycol. Parameter values with an asterisk are taken from dielectric measurements (see Table II).}
	\label{tab:Comp}
\end{table}

%%%%%%%%%%%%%%%%%%%%% begin figure %%%%%%%%%%%%%%%%%%%%%%%%%%%%%%%%%%%%%
\begin{figure}
\hspace{-0cm} \vspace{0cm} \epsfig{file=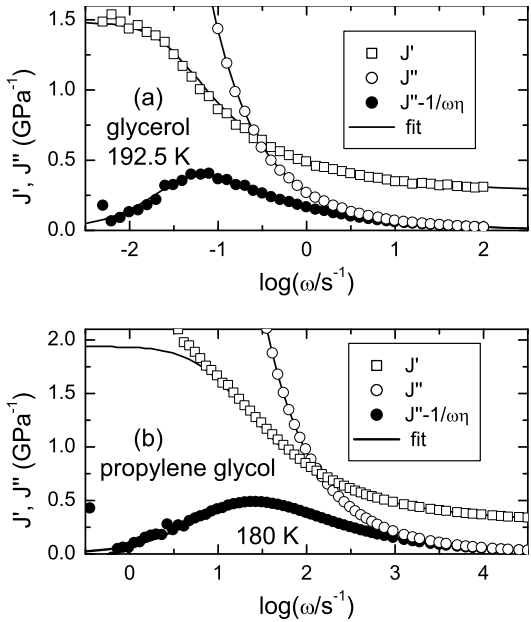,width=7 cm,angle=0} \vspace{0cm} \caption{(a) The glycerol data \cite {donth1} of Fig. 3 (b) transformed to $J(\omega)$ and fitted with a sharp $\tau_c$ (b) Propylene glycol data \cite{maggi} in terms of $J(\omega)$ and fitted with a distribution of $\tau_c$ (parameters see Table I).}
\end{figure}
%%%%%%%%%%%%%%%%%%%%% end figure %%%%%%%%%%%%%%%%%%%%%%%%%%%%%%%%%%%%%%%

Measurements of $G(\omega)$ need to be very accurate to be able to determine a reliable value of $W_r$. In most cases, the five fit parameters $G$, $J_0$, $\eta$, $\beta$ and a sharp $\tau_c$ provide good fits. Fig. 2 shows the two examples vitreous silica \cite{mills} and glycerol \cite{donth1}, which are both describable with a sharp $\tau_c$ within the error bars. Note that in silica the condition $\omega\tau_c=1$ is met at the peak of $G''(\omega)$, while in glycerol it lies a decade below the peak.

In order to attack the problem of the width $W_r$ on the basis of a high-quality measurement of $G(\omega)$, it is necessary to determine first a reliable value of the viscosity \cite{tina} from the low-frequency end of $G''(\omega)\approx\omega\eta$. Having this, one can invert $G(\omega)$ to $J(\omega)$ and subtract the viscosity contribution to look at the end of the retardation processes. In Table I, $\eta$ is given in the form of the Maxwell time $\tau_M=\eta/G$.

%%%%%%%%%%%%%%%%%%%%% begin figure %%%%%%%%%%%%%%%%%%%%%%%%%%%%%%%%%%%%%
\begin{figure}
\hspace{-0cm} \vspace{0cm} \epsfig{file=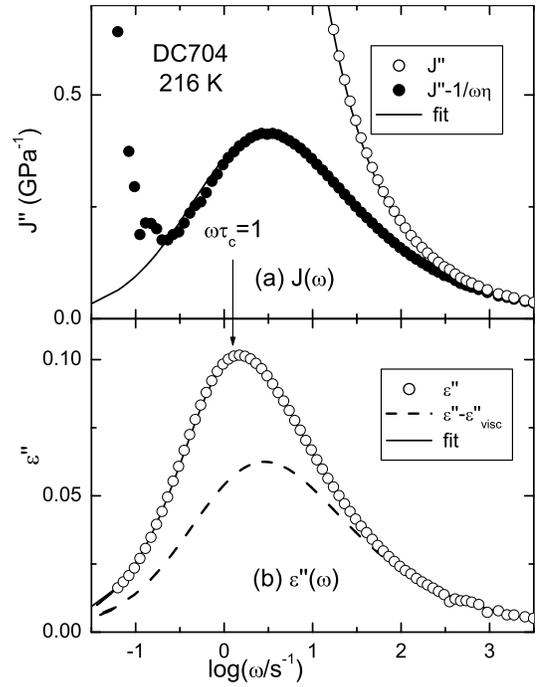,width=7 cm,angle=0} \vspace{0cm} \caption{(a) Fit of dynamical shear data \cite{tina} of DC704 at 216 K (b) Fit of dielectric data \cite{tinan} at the same temperature (parameters see Tables I and II).}
\end{figure}
%%%%%%%%%%%%%%%%%%%%% end figure %%%%%%%%%%%%%%%%%%%%%%%%%%%%%%%%%%%%%%%

Fig. 4 (a) shows the result of this procedure in the pioneering work of Schr\"oter and Donth \cite{donth1} for glycerol. The fit of the present model yields $W_r=0\pm 2$, still compatible with a sharp $\tau_c$. The same remains true for another hydrogen-bonded example, propylene glycol \cite{maggi} in Fig. 4(b), though in this case the introduction of a non-zero width already improves the fit.

If one proceeds to van-der-Waals bonded molecules, one begins to find larger values of $W_r$. The measurements are no longer compatible with a sharp $\tau_c$. The first example 5-PPE, a vacuum pump oil, has $W_r$-values around 3, with an estimated error which is a factor of two smaller (see Table I). These are data from a recent and highly accurate measurement \cite{tina}.

Fig. 5 (a) shows a second example from the same paper \cite{tina}, DC704, another vacuum pump oil. In this case, $W_r=2.6\pm0.9$. Comparing to dielectric data (Fig. 5 (b)) measured in the same cryostat for a sample from the same charge \cite{tinan}, one finds the peak of $\epsilon''(\omega)$ at the condition $\omega\tau_c=1$ fitted to the shear data. Even more, if one fits the dielectric data in the same model, the retarded part of the dielectric response turns out to be identical with the retarded part of the shear response within experimental error.

The two parameters $\tau_c$ and $W_r$ have much more accurate values in the dielectric fits. In the next example of Table I, propylene carbonate \cite{donth2}, the shear fit was done using the value of $W_r$ from dielectric data \cite{schneider} in Table II.

For triphenylethylene, one has shear and dielectric measurements in the same cryostat \cite{niss}. Therefore one can take both $W_r$ and $\tau_c$ from the dielectric data and gets excellent fits for the shear.

The same procedure works perfectly well for the three substances DGEBA \cite{donth2,comez}, propylene glycol \cite{maggi,albena} (Fig. 4) and diethyl phthalate \cite{maggi,albena} (Fig. 7), though in these cases dielectric and shear data were not measured in the same cryostat (see Tables I and II).

\subsection{Dielectrics}

%%%%%%%%%%%%%%%%%%%%% begin figure %%%%%%%%%%%%%%%%%%%%%%%%%%%%%%%%%%%%%
\begin{figure}
\hspace{-0cm} \vspace{0cm} \epsfig{file=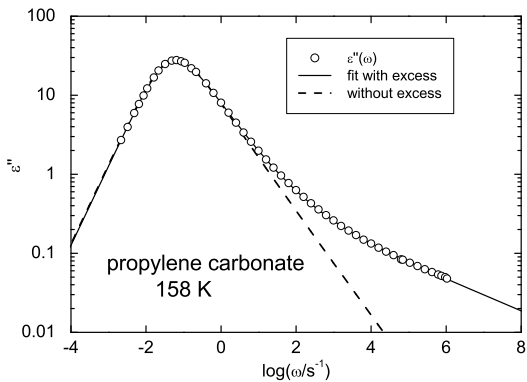,width=7 cm,angle=0} \vspace{0cm} \caption{ Fit of dielectric data \cite{schneider} of propylene carbonate at 158 K. At the peak, the fits with and without excess term are the same (parameters see text and Table II).}
\end{figure}
%%%%%%%%%%%%%%%%%%%%% end figure %%%%%%%%%%%%%%%%%%%%%%%%%%%%%%%%%%%%%%%

%%%%%%%%%%%%%%%%%%%%% begin figure %%%%%%%%%%%%%%%%%%%%%%%%%%%%%%%%%%%%%
\begin{figure}
\hspace{-0cm} \vspace{0cm} \epsfig{file=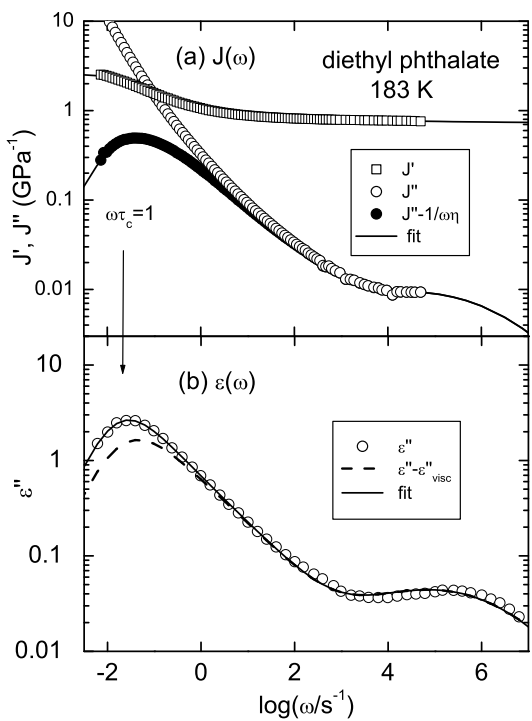,width=7 cm,angle=0} \vspace{0cm} \caption{ Combined fit of (a) shear \cite{maggi} and (b) dielectric data \cite{albena} of diethyl phthalate at 183 K.}
\end{figure}
%%%%%%%%%%%%%%%%%%%%% end figure %%%%%%%%%%%%%%%%%%%%%%%%%%%%%%%%%%%%%%%

%%%%%%%%%%%%%%%%%%%%% begin figure %%%%%%%%%%%%%%%%%%%%%%%%%%%%%%%%%%%%%
\begin{figure}
\hspace{-0cm} \vspace{0cm} \epsfig{file=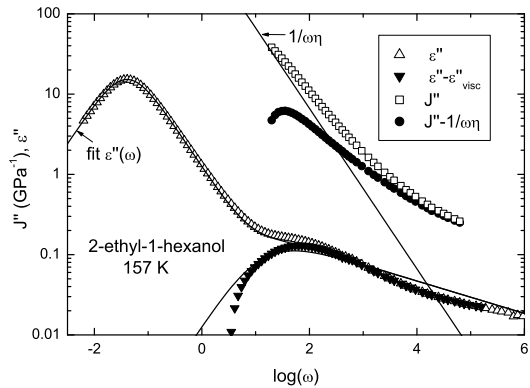,width=7 cm,angle=0} \vspace{0cm} \caption{Comparison of the retardation fractions in shear and dielectrics for a mono-alcohol.}
\end{figure}
%%%%%%%%%%%%%%%%%%%%% end figure %%%%%%%%%%%%%%%%%%%%%%%%%%%%%%%%%%%%%%%

The fit for DC704 in Fig. 5 (b) had the five parameters $\Delta\chi_1=\epsilon_s-\epsilon_\infty$, $\tau_c$, $\beta$, $W_r$ and $W_\eta$ (see Table II). Three of them, $\tau_c$, $\beta$ and $W_r$, were identical within experimental error with those fitted to the shear data in Fig. 5 (a).

But naturally, the accuracy of the three parameters is much better in the dielectric fit. Even more important, the dielectric data supply the information that there is nothing at longer times, an information which is impossible to obtain from the shear data without modeling. The point is that the viscous response is different: in dielectrics, one has a broadened line at $\omega\tau_c=1$, in shear, one has the permanent viscous flow.  

There is some correlation between $W_r$ and $W_\eta$ in the fit of Fig. 5 (b), but nevertheless one gets the clear information that $W_\eta$ is only about half of $W_r$. This is again compatible with the NMR result \cite{bohmer} of a decay of the molecular orientation in many small and a few large angular steps. Obviously, the molecular orientation is not only influenced by the processes within the inner relaxing core, but also by the processes in the neighborhood.

The same findings are obtained from the comparison of shear \cite{tina} and dielectric data \cite{tinan} of 5-PPE (see Tables I and II).

The two cases DC704 and 5-PPE are the only ones in which the shear data are accurate enough to determine not only the average $\tau_c$, but also the width $W_r$ of its distribution.

Since the dielectric measurements \cite{tinan} stretch over a large temperature range, they also provide the information that $W_\eta$ and $W_r$ are temperature-independent within experimental error, an information which is compatible with the time-temperature superposition of the spectra found in many glass formers.

Triphenylethylene \cite{niss} and DGEBA \cite{donth2,comez} are two cases where the $\tau_c$ and $W_r$ values for the shear data can be taken from dielectric fits, though in the second case the measurements were not done in the same cryostat.

\begin{table}[htbp]
	\centering
		\begin{tabular}{|c|c|c|c|c|c|c|}
\hline
substance                  & $T$  & $\Delta\chi_1$     & $\tau_c$           & $\beta$ &$W_\eta$   &  $W_r$   \\
\hline   
                           & $K$  &                    & $s$                &         &           &         \\
\hline                                                   
DC704 \cite{tinan}         & 216  & 0.341              & 0.93               &  0.45   & 1.4  & 2.6$\pm$0.5  \\
5-PPE \cite{tinan}         & 250  & 1.895              & 4.9                &  0.48   & 1.1  & 2.1$\pm$0.4  \\
TPE \cite{niss}            & 258  & 0.046              & 0.53               &  0.51   & 1.7  & 3.0$\pm$0.7  \\
DGEBA \cite{comez}         & 256.3& 7.92               & 11.3               &  0.45   & 1.6  & 4.1$\pm$0.5  \\
\hline
glycerol$^a$ \cite{schneider}  & 190  & 70.2               & 26.8               &  0.59   & 0.9  & 1.9$\pm$0.4  \\
PC$^b$ \cite{schneider}        & 158  & 74.8               & 20.6               &  0.64   & 0.8  & 1.65$\pm$0.4 \\
PG$^c$ \cite{albena}           & 180  & 65.2               & 0.101              &  0.65   & 0.5  & 1.37$\pm$0.4 \\
\hline
DEP$^d$ \cite{albena}          & 183  & 8.15               & 50.7               &  0.43   & 1.2  & 1.73$\pm$0.5 \\
\hline
MONO-1$^f$ \cite{boj} & 157& 31.3               & 0.11               &  0.22   & 0.28 & 1.5          \\
\hline		
		\end{tabular}
	\caption{Fit parameters for dielectric data in nine glass formers (significance of substance names see Table I). $^a$ Glycerol has an excess wing with $f_{1/6}=0.04$. $^b$ Propylene carbonate has an excess wing with $f_{1/6}=0.042$. $^c$ Propylene glycol has an excess wing with $f_{1/6}=0.025$. $^d$ Diethyl phthalate has a secondary peak with $f_\beta=0.017$, $V_\beta$ = 0.24 eV and $\Delta V_\beta$ = 0.4 $V_\beta$. $^f$ Mono-alcohol 2-ethyl-1-hexanol with $\tau_\eta=224\tau_c$ (all other cases in Table II have $\tau_\eta=\tau_c$).}
	\label{tab2:Comp}
\end{table}

In the three examples glycerol \cite{schneider}, propylene glycol \cite{albena} and propylene carbonate \cite{albena} (Fig. 6) one needs to include an excess wing at high frequencies. This was done assuming 
$l(\tau)\propto (\tau/\tau_c)^\beta+f_{1/6}(\tau/\tau_c)^{1/6}$, with the $f_{1/6}$-values of the three substances given in Table II.

Similarly, the example diethyl phthalate \cite{albena} required the inclusion of a $\beta$-peak. The $\beta$-peak was described by a gaussian distribution of barrier heights around an average barrier $V_\beta$ with a full width at half maximum of $\Delta V_\beta$, calculating the relaxation times from $\tau=\tau_0\exp(V/kT)$ with $\tau_0=10^{-13}s$. The resulting normalized relaxation time density is denoted by $L_\beta(\tau)$. The strength of the $\beta$-relaxation is characterized by $f_\beta$, again with $l(\tau)\propto (\tau/\tau_c)^\beta+f_\beta L_\beta(\tau)$.

Fig. 7 (a) and (b) show the fits. Interestingly, $f_\beta$ turned out to be a factor of nearly two higher in the dielectric fit than in the shear one, indicating a different ratio of relaxation strengthes for $\alpha$ and $\beta$-peak.

A very important case for the model of the present paper turns out to be the one of the mono-alcohols \cite{boj}, where the terminal relaxation time $\tau_\eta$ of the viscous mode lies more than a factor of hundred higher than the relaxation time $\tau_c$ of the local structure, thus enabling a separate and unambiguous determination of the two components.

Fig. 8 shows the example of 2-ethyl-1-hexanol with $\tau_\eta=224\tau_c$. Most of the dielectric signal is found in a practically sharp Debye line at $\tau_\eta=24.6$ s. If one subtracts this sharp line from the data, one finds the small retardation contribution shown in the figure. Comparing this to the retardation contribution $J''-1/\omega\eta$ of the shear compliance, one sees that within experimental error both seem to have the same low frequency cutoff. The same result is obtained for 2-butanol, the second mono-alcohol for which one has shear and dielectric data measured in the same cryostat for the same sample \cite{boj}.

This is a very important finding. If dielectrics and shear have the same $\tau_c$, this implies that it is not only the time where a local structure forgets its elastic misfit (this was the starting hypothesis of the present paper), but rather a general stop of all back-and-forth motion, after which any remaining memory must be removed by viscous no-return jumps. In fact, this new hypothesis is the basis of the equations in Section II B.

\section{Discussion and Conclusions}

To understand the results in Table I properly, it is important to realize that $G(\omega)$-data can never determine a reliable cutoff for the back-and-forth jumps of the retardation without additional assumptions.

In the model of this paper, the additional assumption is twofold: (i) one assumes a density of relaxators increasing proportional to $(\tau_r)^\beta$ without end (ii) the relaxators decay with a gaussian distribution of lifetimes. It is this twofold assumption which forces the $G(\omega)$ data to supply an end of the back-and-forth motion at long times. With other assumptions, the data would allow for a happy continuation forever.

The model is supported by the dielectric findings compiled in Table II. They show that one can explain the dielectric data in terms of the same model, with the same lifetime and the same lifetime distribution as the one of the shear fit. The Kohlrausch $\beta$ parameter agrees within the error bars in five of the cases, but differs in the three excess-wing examples glycerol, propylene carbonate and propylene glycol, possibly because the shear data were fitted without excess wing.

The $W_r$-values in Table II vary from 1.37 to 4.1, a much larger variation than the error bars of the determination. One concludes that $W_r$ is not a universal value, but varies from substance to substance. In those substances where one can check, one finds a temperature-independent $W_r$. 

The example diethyl phthalate shows the applicability of the model to glass formers with a secondary relaxation peak.

If one concedes the existence of a viscous part of the dielectric response, it is clear that one needs to subtract this viscous part before being able to say anything about the fraction of the retardation due to the $\beta$-peak. With the model, this is possible. In the case of diethyl phthalate, one finds different contributions of the $\beta$-peak to the dielectric and to the shear retardation.

A strong argument for the validity of the model is supplied by the mono-alcohol fit in Fig. 8. There, the dielectric data \cite{boj} show a clear separation of the retardation response from the terminal viscous one. Within the error bars, the maximum of the dielectric retardation response agrees with the one of the shear compliance. Moreover, eq. (\ref{mono}) of the model describes successfully the giant size of the dielectric viscous response.

Of course, a pragmatical model like the present one can only give limited insight. The physical reason for the persisting rise of the retardation processes with increasing relaxation time, describable with a Kohlrausch $\beta$ close to 1/2, remains hidden. But the model is able to show that such a rise exists up to relaxation times which are longer than the lifetimes of local structures. The model is further able to show that the local structure lifetimes have a distribution width of the order of a decade, a distribution width which is different in different glass formers. 

A companion paper shows that the model is essential for a quantitative understanding of nonlinear dielectric data. From the construction of the model, one can hope that it might also be useful for a quantitative description of aging data.

To summarize, an elementary consideration on the contributions of a single barrier to the retardation and to the flow of an undercooled liquid suggests a continuous description in terms of reversible retardation processes at short times and irreversible viscosity processes at long times. The consideration can be cast into a simple convenient model, assuming a density of relaxators compatible with a Kohlrausch function on the fast side. The relaxators decay with a broad distribution of terminal relaxation times centered around an average terminal relaxation time.

\end{document}